\documentclass[conference, 10pt]{IEEEtran}
\usepackage[utf8]{inputenc} 
\usepackage[T1]{fontenc}
\usepackage{afterpage}
\usepackage{algorithm,algorithmic}
\usepackage{amsmath, bm}
\usepackage{amssymb}
\usepackage{caption}
\usepackage{dsfont}
\usepackage{enumitem}
\usepackage{graphicx}
\usepackage{multirow}
\usepackage[short]{optidef}
\usepackage{subcaption}
\usepackage{todonotes}
\usepackage{url}
\usepackage{bbm}

\newcommand{\bs}[1]{\bm{#1}}
\newcommand{\mc}{\mathcal}

\newcommand{\herm}[1]{\ensuremath{#1^{\textsf{H}}}}
\newcommand{\tran}[1]{\ensuremath{#1^{\textsf{T}}}}
\newcommand{\mat}[1]{\ensuremath{\bs{#1}}}
\newcommand{\vect}[1]{\ensuremath{\bs{#1}}}
\newcommand{\dom}[1]{\ensuremath{\mc{#1}}}
\newcommand{\ind}[1]{\mathds{1}_{\left\lbrace #1 \right\rbrace}}

\newcommand{\Mm}{\ensuremath{\mat{M}}}
\newcommand{\Hm}{\ensuremath{\mat{H}}}

\newcommand{\Zm}{\ensuremath{\mat{\zeta}}}

\newcommand{\wv}{\ensuremath{\vect{w}}}
\newcommand{\xv}{\ensuremath{\vect{x}}}
\newcommand{\hv}{\ensuremath{\vect{\zeta}}}
\newcommand{\yv}{\ensuremath{\vect{y}}}

\newcommand{\nv}{\ensuremath{\vect{n}}}

\newcommand{\dW}{\ensuremath{\dom{W}}}
\newcommand{\dX}{\ensuremath{\dom{X}}}

\newcommand{\fenc}[1]{\ensuremath{\phi\left(#1\right)}}

\newcommand{\fmap}[1]{\ensuremath{\phi^{(map)}\left(#1\right)}}
\newcommand{\pdf}[2]{\ensuremath{f_{#1}\left(#2\right)}}
\newcommand{\Esp}[2]{\mathds{E}_{#2}\left[ #1 \right]}
\newcommand{\proba}[2]{\ensuremath{P_{#1}\left(#2\right)}}
\newcommand{\fpost}[4]{\pdf{#1|#2}{#3|#4}}
\newcommand{\Ppost}[4]{\proba{#1|#2}{#3|#4}}
\newcommand{\sigmoid}[1]{\ensuremath{\texttt{sigmoid}\left(#1\right)}}

\title{Joint Constellation Shaping Using Gradient Descent Approach for MU-MIMO Broadcast Channel}
\author{\IEEEauthorblockN{Maxime Vaillant}
\IEEEauthorblockA{\textit{Inria, INSA Lyon, CITI, UR3720} \\
69621 Villeurbanne, France \\
maxime.vaillant@inria.fr}
\and
\IEEEauthorblockN{Alix Jeannerot}
\IEEEauthorblockA{\textit{INSA Lyon, Inria, CITI, UR3720} \\
69621 Villeurbanne, France \\
alix.jeannerot@inria.fr}

\and
\IEEEauthorblockN{Jean-Marie Gorce}
\IEEEauthorblockA{\textit{INSA Lyon, Inria, CITI, UR3720} \\
69621 Villeurbanne, France \\
jean-marie.gorce@insa-lyon.fr}
}

\begin{document}

\maketitle

\begin{abstract}
In this paper, we introduce a learning-based approach to optimize a joint constellation for a multi-user MIMO broadcast channel ($T$ Tx antennas, $K$ users, each with $R$ Rx antennas), with perfect channel knowledge. The aim of the optimizer (MAX-MIN) is to maximize the minimum mutual information between the  transmitter and each receiver, under a sum-power constraint. The proposed optimization method do neither impose the transmitter to use superposition coding (SC) or any other linear precoding, nor to use successive interference cancellation (SIC) at the receiver. Instead, the approach designs a joint constellation, optimized such that its projection into the subspace of each receiver $k$, maximizes the minimum mutual information $I(W_k;Y_k)$ between each transmitted binary input $W_k$ and the output signal at the intended receiver $Y_k$.
The rates obtained by our method are compared to those achieved with linear precoders.  
\end{abstract}
\begin{IEEEkeywords}
Non Degraded Broadcast Channel, Joint Constellation Design, Stochastic Gradient Descent, MU-MIMO
\end{IEEEkeywords}

\section{Introduction}


Power-domain non-orthogonal multiple access (PD-NOMA) is a promising technique relying on superposition coding (SC) to increase the capacity of cellular networks. When a base station (BS) has a single antenna, the corresponding channel is known as the degraded broadcast channel, and SC is capacity achieving  \cite{el2011network}.
When the BS has multiple antennas, the corresponding broadcast channel model (known as the MU-MIMO) is non degraded and the capacity region is not given by a single letter expression \cite{weingarten2006capacity}. Dirty paper coding (DPC) with time-sharing is the only known method allowing to achieve the capacity limits but its complexity is prohibitive \cite{weingarten2006capacity}.
It is worth mentionning that PD-NOMA has been adapted to MU-MIMO, with sub-optimal techniques  such as using predefined beams or successive zero-forcing encoders \cite{liu2018multiple,clerckx2013mimo,clerckx2021noma,vaezi2019non}.
In \cite{clerckx2021noma} PD-NOMA is compared to standard Multi User Linear Precoding (MU-LP), e.g. zero-forcing, and it is shown that in some conditions, they have in practice similar performance. 
Beyond PD-NOMA, rate splitting multiple access (RSMA) overcomes some of the limitations of PD-NOMA \cite{joudeh2016robust}. With RSMA messages are split between private and common parts. The private part is decoded by the intended receiver only, while the common part is decoded by everyone. This promising technique investigated for 6G allows to better exploit the degraded broadcast channel but its complexity is high. 


An alternative to leverage on the full capacity region of the degraded broadcast channel, is to design a joint constellation. Indeed, in the aforementioned techniques (PD-NOMA or RSMA), each stream is assumed to span a whole subspace, generating continuous interference to each other. 
However, for small constellation sizes, the input signals do not span a continuous space but use  discrete positions only. Such consideration has already been considered to optimize structured codes in multi-user settings (see for instance \cite{padakandla2018achievable}).
If constellation design for the broadcast channel has been studied in several papers (\textit{e.g.} \cite{albergeConstellationDesignDeep2018, cejudoPowerAllocationConstellation2017, leeMultiuserSuperpositionTransmission2016}), these works designed superposed constellations (instead of joint constellation) at the transmitter \cite{albergeConstellationDesignDeep2018} and assumed that the use of successive interference cancellation (SIC) at the receiver.

Our work differs from two aspects. First, we do not consider a linear combination of per-user constellations, but we optimize a joint constellation for the multiple users. Second, we do not assume any specific decoder, but we instead maximize pairwise mutual information .
More precisely, the contributions of this paper follow:
\begin{enumerate}
    \item In contrast to the commonly used MU-LP, a joint constellation is designed for simultaneously active users. 
    \item  Such a constellation maximizes the minimal mutual information between the BS and each user, ensuring a better fairness compared to the state of the art method. 
    \item The optimization is agnostic to the decoding technique and maximizes the mutual information. 
    \item Numerical simulations are provided supporting the benefit of our method, compared to standard MU-LP methods in the case where the number of receive antennas is 1. 
\end{enumerate}

Our paper is organized as follows. Section \ref{sec:model} describes the system model and associated assumptions, while Section \ref{sec:jcd} introduces the joint constellation design problem. A stochastic gradient based algorithm is proposed in Section \ref{sec:algo}, allowing to find an (at least locally) optimal constellation. The proposed algorithm is evaluated in Section \ref{sec:num} on small simulated scenarios. 

\section{System Model}
\label{sec:model}
\subsection{Channel model}
A MU-MIMO broadcast channel with a BS equipped with $T$ transmit antennas aiming to transmit simultaneously data symbols to $K$ users, each equipped with $R$ receive antennas. We consider a mapping problem in a one channel use.  Let $W_k$ be the random variable (r.v.) representing the $k^{th}$ user symbol, taking values  $w_k\in \mathcal{W}_k$, where $\mathcal{W}_k$ is the codebook of user $k$. $\vect{W}=[W_1\ldots W_K]$ refers to the random vector aggregating all input r.v. The symbols $W_k$ are drawn independently and uniformly at random: $\mathrm{Pr}(\vect{W}=\wv)=\prod_{k=1}^K\frac{1}{|\mathcal{W}_k|}$. When necessary, we will use the notation $w_k^n$ to denote the $n$-th realization of the random variable $W_k$. However, for the sake of clarity, the subscript $n$ is omitted when not explicitly required. For readability purpose we denote $\sum_{\wv}$ as the summation over $\wv \in \dW_1 \times \dots \times \dW_k$.

The BS implements a deterministic encoder : 
\begin{equation}
\label{eq:encoder}
    \phi:\bs{\dW} =\dW_1 \times \dots \times \dW_K\to\mathbb{C}^T,
\end{equation}
which, maps any $\wv$ to a vector $\xv=\fenc{\wv}$ of length $T$. 
The set of all possible vectors $\xv$  denoted as $\dom{X}$, constitutes the joint constellation. We denote by $\dom{X}_k(w_k)$ the subset of vectors that carry the specific message $w_k$ for user $k$.

The vector $\xv$ is transmitted over the $T$ transmit antennas to the users. The channel between the $T$ transmit antennas and the $R$ receive antennas of user $k$ is a matrix denoted  $\Hm_k$. This matrix is rewritten as: $\vect{H}_k = \sqrt{g_k} \cdot \Zm_k$ where $g_k$ is the square Frobenius norm of the channel matrix and $\Zm_k \in \mathbb{C}^{T \times R}$ is the normalized channel matrix. 
Each user observes a projection $\yv_k \in \mathbb{C}^{R}$ of the input vector $\xv$, passing through the channel:
\begin{equation}
    \label{eq:y_k}
    \yv_k = \herm{\Zm_k} \cdot \xv + \bm{\nu}_k, 
\end{equation}
with $\bm{\nu}_k \sim \mc{CN}(0,\sigma^2_k\cdot I_R)$ with $\sigma_k^2=\sigma^2/g_k$ and $I_R$ the $R\times R$ identity matrix. Additionally, by defining $\Zm = \begin{bmatrix} \Zm_1 & \dots & \Zm_K \end{bmatrix} \in \mathbb{C}^{T \times (K \cdot R)}$ and $\yv \in \mathbb{C}^{K\cdot R}$, the whole system is described as:
\begin{equation}
    \yv=\herm{\Zm} \cdot \xv +\nv,
\end{equation}
with $\nv = \begin{bmatrix} \nu_{1,1} & \dots & \nu_{1,R} & \dots & \nu_{K,1} & \dots & \nu_{K,R} \end{bmatrix} \in \mathbb{C}^{K \cdot R}$, and $\yv = \begin{bmatrix} y_{1,1} & \dots & y_{1,R} & \dots & y_{K,1} & \dots & y_{K,R} \end{bmatrix}$. 
Throughout this paper, we assume that the base station has perfect knowledge of the channel (perfect CSIT).


\subsection{Linear methods (MU-LP)}
 \label{sec:linear}
MU-LP are based on the design of independent constellations with perfect CSIT. The encoding function defined in \eqref{eq:encoder} is split in two parts:
\begin{enumerate}[label=(\roman*)]
    \item A set of mapping functions of the form: $\fmap{\wv} = [\fmap{w_1},\ldots, \fmap{w_K}]$ For instance the BPSK mapping function is $\fmap{0} = 1$ and $\fmap{1} = -1$
    \item An encoding matrix matched to the channel. 
\end{enumerate} 
The encoding function $\phi$ linking symbols $\wv$ to constellation points $\xv$ is thus given by:
\begin{equation}
    \fenc{\wv} = \Mm \cdot \fmap{\wv},
\end{equation}
where \Mm\ is the encoding matrix. To ensure power constraints for these methods, we project the resulting constellation onto a constraint space that will be further defined.
To ease the description of these methods, we consider in the remaining of this section that $R=1$.
\subsubsection{Matched filter} \label{matched_filter}
The matched filter does not consider interference and maximize the signal-to-noise-ratio (SNR). The corresponding encoding matrix is:
\begin{equation}
    \Mm_{m} = \Zm.
\end{equation}
\subsubsection{Zero-forcing filter} \label{zero_forcing_filter}
The zero-forcing (ZF) encoder aims to cancel out all interference caused by the channel using the Moore-Penrose pseudo inverse. Its encoding matrix is:
\begin{equation}
    \Mm_{zf}=\Zm \cdot \left(\herm{\Zm}\cdot \Zm\right)^{-1}.
\end{equation}
The ZF encoder leads to interference free signals at the receivers, but may require in turn a high power at the transmitter especially when the channel vectors are close to each other. 


\subsubsection{MMSE (Wiener) filter} \label{mmse_filter}

The minimum mean square error (MMSE) filtering consists of minimizing the noise in reception. As indicated in \cite{goutay2021machine}, the filter that maximizes the SNR is given by:
\begin{equation}
    \Mm_{mmse}=\Zm \cdot \left(\herm{\Zm}\cdot \Zm + \bs{\Sigma} \right)^{-1},\quad \bs{\Sigma} = \textit{diag}(\sigma_i^2)
\end{equation}

\section{Joint Constellation Design} \label{sec:jcd}
In contrast to the MU-LP methods aforementioned, this paper explores the design of a joint constellation that directly maps any input vector $\wv$ to a constellation point $X(\wv)$ according to \eqref{eq:encoder}.
The proposed mapping function is designed to maximize pairwise mutual information as described now.

\subsection{MAX-MIN encoder}
The proposed MAX-MIN encoder aims at maximizing the minimal mutual information between the BS and each user, but the approach could be easily generalized to any other combination of individual losses. 
Given a cross-entropy loss $\mc{L}_k$ (details are provided later), the MAX-MIN optimal solution is a set of constellation points $\dX=\left\{X(\wv)=\fenc{\wv}; \forall \wv\right\}$ that verify:
\begin{argmini}
    {\dX}{\max_k \left( \mc{L}_k(\dX) \right)}{}{}
    \addConstraint{\frac{\sum_{\wv}||X(\wv)||^2}{|\bs{\dW}|}}{\leq P_m}{}
    \addConstraint{||x_t||^2} {\leq P_c\quad} {\forall t \in T, \quad \forall \xv \in \dX},
\end{argmini}
where the first constraint stands for an average power constraint $P_m$ and the second constraint ensures that the transmit power on any antenna does not exceed a threshold $P_c$. We denote $\Pi_{\dom{P}}$ as the constraint space. 

The following sub-sections allow to define the cross-entropy loss that minimizes the mutual information and is appropriate to derive a stochastic gradient descent algorithm. 

\subsection{Log-likelihood ratio (LLR) and mutual information}
Given an observation $\yv_k$ and a symbol $w_k$ the log-likelihood ratio (LLR) is given by:
\begin{equation}
    \label{eq:LLRsref}
     LLR(w_k|\yv_k)
    = \ln\left( \frac{\sum_{\tilde{\wv}; \tilde{w}_k=w_k} \fpost{\bs{Y}_k}{\widetilde{\bs{W}}}{\yv_k}{\tilde{\wv}}}{\sum_{\tilde{\wv}; \tilde{w}_k \neq w_k} \fpost{\bs{Y}_k}{\widetilde{\bs{W}}}{\yv_k}{\tilde{\wv}}} \right).
\end{equation}
This LLR is relevant in communication systems as it is the usual input for soft-decision based decoders. 
Given the prior distribution, the LLR is proportional to the Maximum A Posteriori (MAP) ratio, noted: 
\begin{equation}
    \label{eq:MAP}
    \begin{split}
    LLR_{0}(w_k|\yv_k) &= \ln\left( \frac{\Ppost{W_k}{\bs{Y}_k}{w_k}{\yv_k}}{\sum_{w_i\neq w_k}\Ppost{W_k}{\bs{Y}_k}{w_i}{\yv_k}} \right).
    \end{split}
\end{equation}
It is worth noting that the posterior probability can be obtained from the $LLR _0$ with the sigmoid function $\sigmoid{x}=\frac{1}{1+e^{-x}}$, with:
\begin{equation}
    \label{eq:sigmoidLLR}
   \Ppost{W_k}{\yv_k}{w_k}{\yv_k}= \sigmoid{LLR_0(w_k|\yv_k)}.
\end{equation}
Reminding that the mutual information between $\bs{Y}_k$ and $W_k$ is :
\begin{equation}        \label{eq:mutual_information}
        I(W_k;\bs{Y}_k)= H(W_k)- H(W_k|\bs{Y}_k), 
\end{equation}
where $H(W_k)=\log_2(|\mathcal{W}_k|)$ is constant in our setup since the input distribution is fxed. Further the equivocation is given by: 
\begin{equation}
\label{eq:equivoc}
    \begin{split}
    H(W_k|\bs{Y}_k) &= \Esp{ - \log_2\left(\Ppost{W_k}{\bs{Y}_k}{w_k}{\yv_k}\right)  }{\bs{Y}_kW_k},
    \end{split}
\end{equation}
And from \eqref{eq:equivoc}, the equivocation depends directly on the average $LLR_0$ defined in  \eqref{eq:MAP}:
\begin{equation}
    \label{eq:equivocation}
    H(W_k|\bs{Y}_k) = \Esp{-\log_2\left(\sigmoid{LLR_0(w_k|\yv_k)}\right)}{\bs{Y}_kW_k}.
\end{equation}
This let us to conclude that maximizing mutual information under uniform input distributions relies on maximizing the averaged $LLR_0$, used below to define the loss function. 

\subsection{Cross-entropy loss}
In the context of machine learning, cross-entropy is used to optimize the parameters of a model. It compares the labels of the training data to the decisions made at the output of a system. We propose to optimize the constellation points by evaluating the quality of simulated received data with their respective posterior probability distributions. The cross-entropy loss associated to a batch of $N$ simulations, is expressed as:
\begin{equation}
    \label{eq:cross-entropy}
    \begin{split}
        \mc{L}_k &=- \frac{1}{N} \sum_{n=1}^N \sum_{w \in \dom{W}_k} \ind{W_k^n=w}\cdot \log_2\left(P_{{W_k}|\bs{Y}_k}(W_k^n=w|\yv_k^n)\right)  \\
        &\underset{N\rightarrow\infty}{\longrightarrow} \Esp{-\log_2\left(P_{{W}_k|\bs{Y}_k}({w}_k|\yv_k)\right)}{W_k \bs{Y}_k} = H(W_k|\bs{Y}_k).
    \end{split}
\end{equation}


The limit when $N\rightarrow\infty$ shows that the loss $\mc{L}_k$ converges to the equivocation (\ref{eq:equivoc}) which is directly linked to the mutual information (\ref{eq:mutual_information}). 
We cannot guarantee that a gradient descent will reach the optimal point, but this proves that asymptotically, the minimum of the Loss function corresponds to the mutual information maximum.
For simplification, we introduce the notation:
\begin{equation}
    \label{eq:lk}
    \mc{L}_k(\dX;w_k^n;\yv_k^n) = - \log_2 \left( \sigmoid{LLR_0(\dX;w_k^n|\yv_k^n)} \right),
\end{equation}
which represents the loss on user $k$ for a given sample $n$. 
Using (\ref{eq:MAP}) and Gaussian noise statistics, we have:
\begin{equation}
    \label{eq:llr_0}
    LLR_0(\dX;w_k^n|\yv_k^n) = \ln \left( \frac{\sum_{\xv \in \dom{X}_k(w_k^n)}e^{-d_k(\xv)}}{\sum_{\xv \notin \dom{X}_k(w_k^n)}e^{-d_k(\xv)}} \right),
\end{equation}
where the exponent in the Gaussian noise distribution formula is $d_k(\xv) = \frac{||\yv_k - \herm{\Zm_k}\cdot \xv ||^2}{2\sigma_k^2}$. The aim of \eqref{eq:llr_0} is to exhibit the loss with respect to the reference and observed constellation points.

The overall loss (\ref{eq:cross-entropy}) can then be expressed as a function of $\dX$:
\begin{equation}
    \label{eq:overall_loss}
    \mc{L}_k(\dX) = \frac{1}{N} \sum_{n=1}^N \sum_{w \in \dom{W}_k} \ind{w_k^n=w}\cdot \mc{L}_k(\dX;w_k^n;\yv_k^n).
\end{equation}

The function $\mc{L}_k(\dX)$ is differentiable in $\dX$ which makes it suitable for optimization methods based on stochastic gradient that will lead to local stationary point as $\mc{L}_k(\dX)$ is non-convex in $\dX$. 

\section{Proposed Algorithms}\label{sec:algo}
In this section, we present the algorithm employed to optimize the previously defined loss function. The optimization algorithm Adam \cite{kingma2014adam}, has been chosen, a variant of Stochastic Gradient Descent (SGD) with momentum. We aim to directly optimize the position of each point of the constellation without restricting it to any MU-LP structure. 
\begin{algorithm}[H]
    \caption{Constellation optimization algorithm using projected stochastic gradient descent}
    \label{alg:psgd}
    \begin{algorithmic}[1]
        \STATE \label{alg:init} Initialize random $\dX$
        \WHILE{not converged} \label{alg:loop}
        \STATE Generate $\{\wv^1, \dots, \wv^N\}$ \label{alg:code_words}
        \FOR{$k \in K$} \label{alg:iter_users}
        \STATE Simulate $\{\yv_k^1, \dots, \yv_k^N\}$ according to (\ref{eq:y_k}) \label{alg:y_k_simulation}
        \STATE $\mc{L}_k(\dX) \gets \frac{1}{N} \cdot \sum_{n=1}^N{\mc{L}_k(\dX;w_k^n;\yv_k^n)}$ \label{alg:user_loss}
        \ENDFOR \label{alg:end_iter_users}
        \STATE $ \mc{L}(\dX) \gets \text{max}(\mc{L}_1(\dX), \dots, \mc{L}_K(\dX))$ \label{alg:losses_reduction}
        \STATE $\dX \gets \Pi_{\dom{P}} [\texttt{ADAM}(\dX, \eta, \mc{L})]$ \label{alg:gradient_step}
    \ENDWHILE \label{alg:end_loop}
    \end{algorithmic}
\end{algorithm}
The proposed iterative algorithm aims to optimize the constellation design while ensuring fairness across all users. The steps of the algorithm follow: 
\begin{itemize}
    \item \textbf{Initialization} (\ref{alg:init}): A random constellation \dX\ is chosen.
    \item \textbf{Generation of codewords} (\ref{alg:code_words}): $N$ random codewords representing $N$ independent realizations of $\vect{W}$ are drawn.
    \item \textbf{Simulation of observations} (\ref{alg:y_k_simulation}): For each vector, $N$ observations are simulated according to the Gaussian channel.
    \item \textbf{Computation of user loss} (\ref{alg:user_loss}): The average individual loss is computed for each user $k$.
    \item \textbf{Reduction Function} (\ref{alg:losses_reduction}): After computing the loss of all users, a reduction function $\mathbb{R}^K\to\mathbb{R}$ is applied, that is the MIN-MAX function. 
    \item \textbf{Gradient Step with Projection} (\ref{alg:gradient_step}): One step of the Adam\footnote{we use the Adam algorithm as implemented in the TensorFlow library} algorithm with step size $\eta$ is performed. The output  is then projected into $\Pi_p$, with a normalization (to ensure $P=P_m$) and a projection to the closest point (to ensure each $p_k\leq P_c$).
\end{itemize}

This algorithm runs until a convergence criterion is met or when the number of iteration reaches a threshold. 


\section{Numerical Results}
\label{sec:num}
The proposed joint constellation design algorithm is compared to MU-LP solutions. Reminding that our primary objective is to maximize the minimum mutual information between the BS and each user, we evaluate the mutual information as a function of the signal-to-noise ratio (SNR), defined by:
\begin{equation}
    \mathrm{SNR}_{dB} = 10 \log_{10}\left( P_m/\sigma_k^2\right).
\end{equation}
In this work, we focus only on the design of the constellation, we thus assume that the encoder (from bits to symbols $W_k$) and the decoder (from received points $\bs{Y}_k$ to estimated bits) are optimal, meaning that the natural metric to consider is the mutual information between $W_k$ and $\bs{Y}_k$ which represents an upperbound on the Bit Error Rate (BER).
In all simulations, we only consider single-antenna receivers ($R=1$) to compare with MU-LP encoders even if our algorithm works for any $R \geq 1$. The hyper-parameters $\eta$ and $N$ (samples per iteration) are respectively set to $0.1$ and $10000$, and we fix $P_c = 4$ and $P_m=1$. This implies that the Peak to Average Power Ratio (PAPR) is at most 4. Alg. \ref{alg:psgd} is run for 100 iterations.
The code\footnote{Available at \url{https://gitlab.inria.fr/maracas/publications/broadcast-mu-mimo-joint-constellation-learning}} of the simulations is developed using the Sionna \cite{sionna} package. In these simulations, we focuse on small per-user modulations, but it is important to note that $1$ bit for $10$ users could also be $2$ bits for $5$ users. Essentially, we can process 10 bits across different users.

The first scenarios are for illustrative purpose, with $T=2$ transmit antennas and $K=2$ users with real channels, allowing to plot the constellations. Subsequently, we present a scenario involving complex channels with $T=4$ and $K=10$. 

\subsection{Real Channels}

Given real channels, the constellation over the 2 antennas are plot on, allowing to compare the MU-LP and learned approaches. Two distinct scenarios are proposed :
\begin{itemize}
    \item \textbf{Scenario 1}: $T=2$ transmit antennas, $K=2$ users with a SNR of $6$ and $8$ dB, respectively and non-orthogonal channels: $\hv_1 = \tran{\begin{bmatrix} 1 & 0 \end{bmatrix}}$, $\hv_2 = \tran{\begin{bmatrix} \frac{\sqrt{2}}{2} & \frac{\sqrt{2}}{2} \end{bmatrix}}$.
    \item \textbf{Scenario 2}: same as Scenario 1 but with colinear channels: $\hv_1 = \tran{\begin{bmatrix} \frac{\sqrt{2}}{2} & \frac{\sqrt{2}}{2} \end{bmatrix}}$, $\hv_2 = \tran{\begin{bmatrix} \frac{\sqrt{2}}{2} & \frac{\sqrt{2}}{2} \end{bmatrix}}$.
\end{itemize}
In both scenario the BS sends a single bit to each user: $\dW_1=\{0,1\}$ and $\dW_2=\{0,1\}$.
\begin{figure}[ht]
    \centering
    \includegraphics[width=0.38\textwidth]{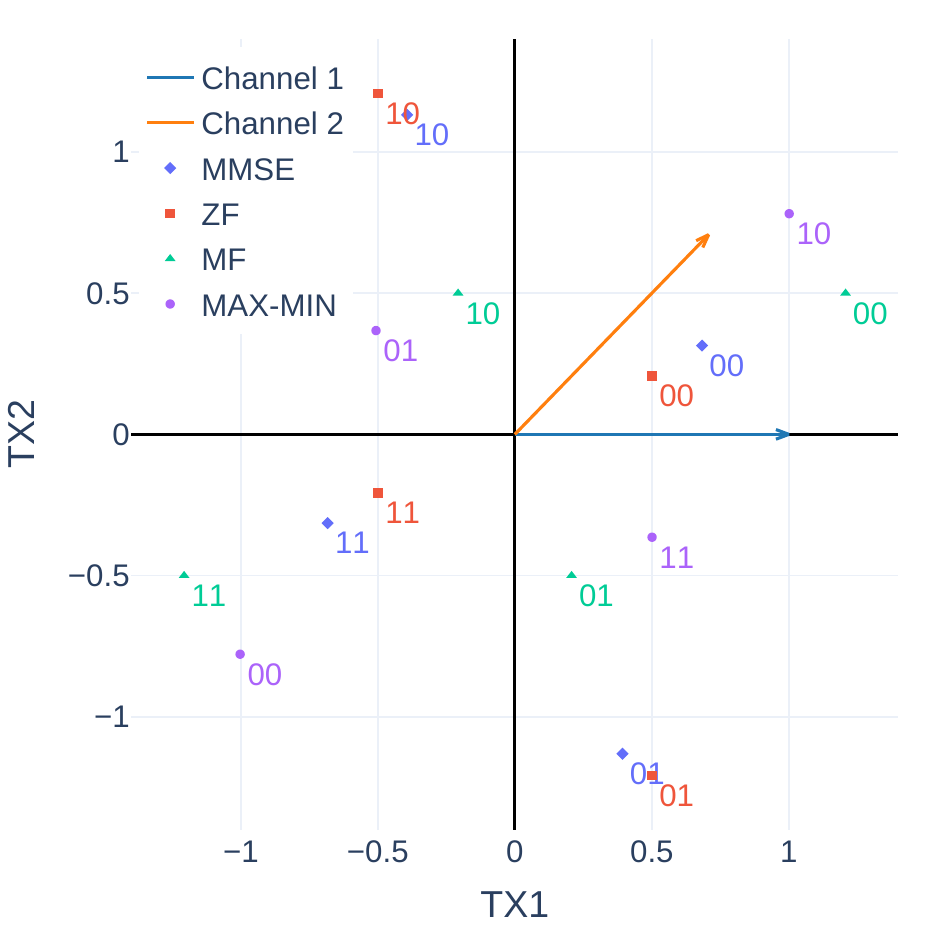}
    \caption{Constellation of the different encoders with non-orthogonal channels}
    \label{fig:scenario_a}
\end{figure}
\begin{figure}[ht]
    \centering
    \includegraphics[width=0.38\textwidth]{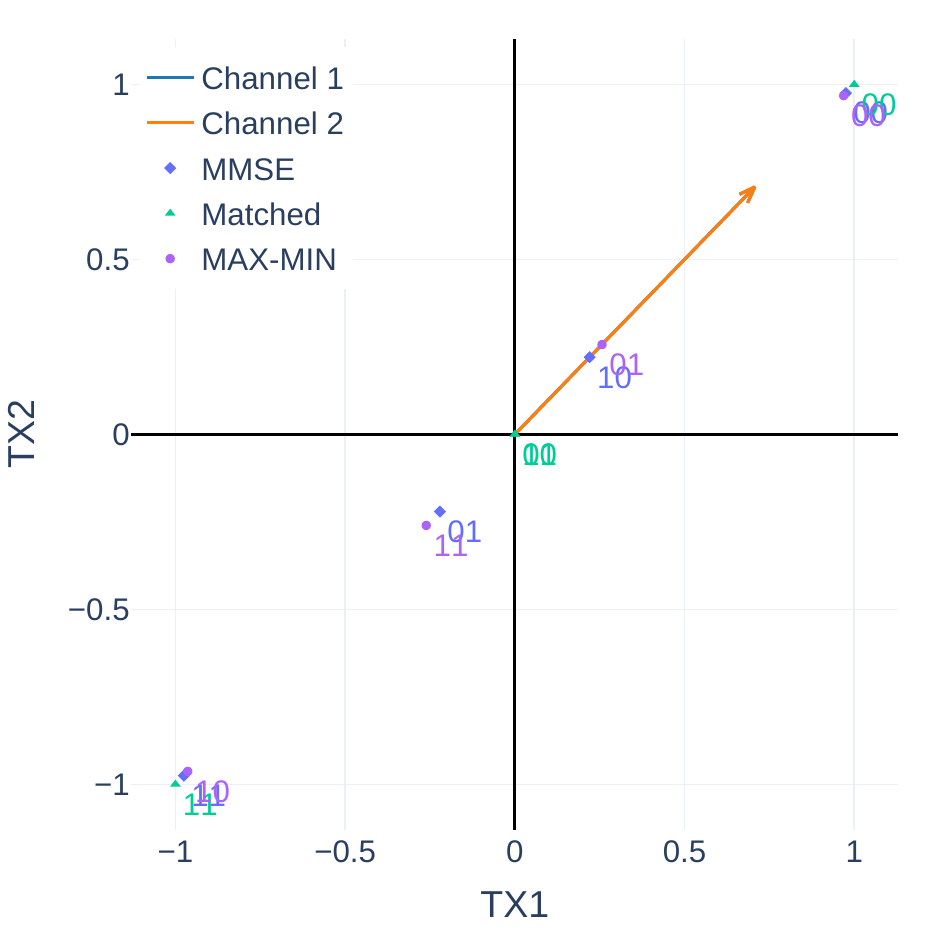}
    \caption{Constellation of the different encoders with colinear channels}
    \label{fig:scenario_b}
\end{figure}
\begin{table}[h]
    \centering
    \caption{Scenario results}
    \begin{tabular}{|c|c|c|c|c|c|}
        \hline
        \textbf{Scenario} & \textbf{Encoders} & $I_1$ & $I_2$ & $\min_k (I_k)$ & $\text{mean}_k(I_k)$\\
        \hline
        \multirow{4}{*}{1} 
        & MMSE & 0.50 & 0.76 & 0.50 & 0.63 \\
        & ZF & 0.48 & 0.64 & 0.48 & 0.56 \\
        & Matched & 0.49 & 0.55 & 0.49 & 0.52\\
        & MAX-MIN & 0.69 & 0.70 & \textbf{0.69} & \textbf{0.69}\\
        \hline
        \multirow{4}{*}{2} 
        & MMSE & 0.28 & 0.66 & 0.29 & 0.47 \\
        & ZF & - & - & - & - \\
        & Matched & 0.39 & 0.44 & 0.39 & 0.42 \\
        & MAX-MIN & 0.63 & 0.63 & \textbf{0.63} & \textbf{0.63} \\
        \hline
    \end{tabular}
    \label{tab:scenario_results}
\end{table}
The resulting constellations are presented in Fig. \ref{fig:scenario_a} and Fig.\ref{fig:scenario_b} where for each point the first and second indicated bits are transmitted to user 1 and user 2 respectively. $\hv_1$ and $\hv_2$ are represented by pink and brown arrow. We observe on Fig. \ref{fig:scenario_a}, that MU-LP approaches yield symmetrical constellations, where each constellation point associated to $b_0b_1$ is centrally symmetrical to the point associated to $\bar{b}_0\bar{b}_1$. In contrast, the learned approach does not exhibit this symmetry. More interestingly, the projection of the constellation points onto each user's observation line shows that th 0s and 1s are split in two decision regions, as usual. On the opposite, the learned approach results in a different pattern: the projection of user 1 reveals a central cluster of 0s surrounded by two clusters of 1s at both extremities. This unique clustering behavior exhibits the kind of optimization such approach can provide and allowing to improve the mutual information, as summarized in Table \ref{tab:scenario_results}. 
On the second scenario, the proposed method learned a joint constellation by adapting the distance between the different points and almost achieve the equal rate capacity given by SC (its value is $0.67$ per user, according to \cite{el2011network}).
Notably, while  our proposed approach achieves fairness across all users, the MMSE approach sacrifices the performance of user 1, with the highest noise level. In addition our approach maintains a strong average performance. Note that in scenario 2, the ZF encoder is not able to cancel interference as the rank of the channel matrix is one, making $\herm{\Hm}\Hm$ not invertible.

\subsection{Complex Channels}
The simulations are now extended to complex channels with more receivers. We focus on a scenario with $K = 10$ users and $T = 4$ transmitting antennas. For each round, an average SNR $s$ (dB) where $s\in[-5,15]$, is fixed, and the SNR $\gamma_k$ of each user is drawn once from $\mathcal{N}(s,1)$ .
We conduct $20$ independent experiments, where each follows: \begin{enumerate}
    \item For each user a random channel vector is drawn from $\mathcal{CN}(0,I_T)$ and is then normalized,
    \item For each average SNR level $s$:
    \begin{itemize}
        \item MU-LP reference constellations are computed according to \ref{sec:linear} and are projected into $\Pi_p$,    
        \item $10$ constellations are drawn randomly and Alg.\ref{alg:psgd} is run on each. The best resulting constellation is recorded. 
    \end{itemize}    
    \item The average and minimum  mutual information values are computed for each constellation. 
\end{enumerate}
\begin{figure}[ht]
    \centering
    \includegraphics[width=0.5\textwidth]{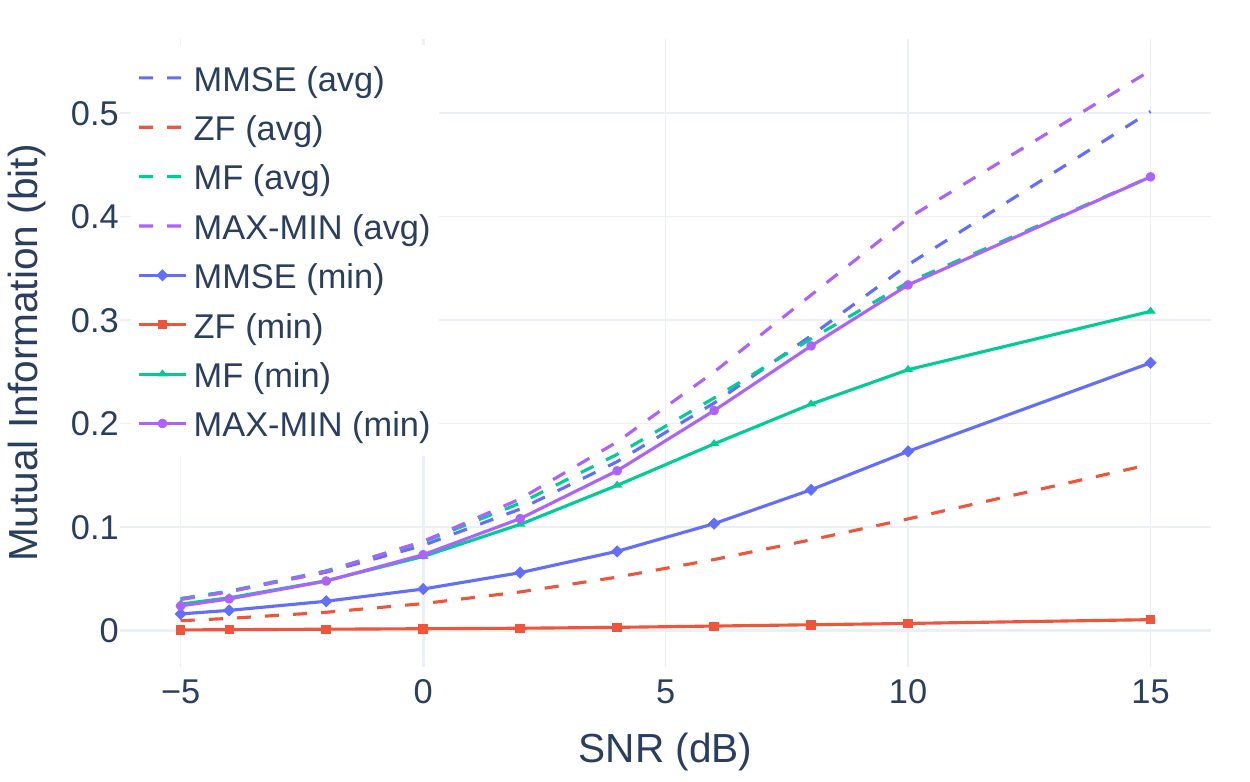}
    \caption{Mutual information on users for the different encoders under varying SNR}
    \label{fig:waterfall_plot}
\end{figure}

On Fig.\ref{fig:waterfall_plot}, plain lines show the minimum mutual information among the users. For all SNR values, the learned constellation outperforms MU-LP constellations, except for the matched encoder at low SNR. 
It is worth noting that the ZF encoder (red curve) does not cancel all interference due to power constraints imposed on the constellation. 
On Fig.\ref{fig:waterfall_plot} dashed lines illustrate the mean of the mutual information on all users. It can be seen that our method outperforms MU-LP in all SNR regimes, while improving the min value up to (42\%).

\section{Conclusion}
In this paper, we proposed an approach to optimize a joint constellation for the MU-MIMO broadcast channel, with a stochastic gradient descent maximizing the minimal mutual information between the BS and each user. By leveraging on a learned approach, we obtained constellations that are not feasible with traditional linear methods. This proves the higher flexibility of the learned technique, especially for non linear optimization (e.g. Min-Max instead of sum-rate) with a significant performance improvement. 

However, this work has some limitations, in particular the exponential growth of the constellation size ($2^K$ points) restricts the value of $K$ (in our case, $10$). Moreover, the proposed solution has to be computed each time the channel changes. Generalization to a more efficient and scalable method, possibly with a neural network, is left as future work.


\section*{Acknowledgment}

This his work was funded by the French Government through the “Plan de Relance” and “Programme d’investissements d’avenir"

\bibliographystyle{IEEEtran}
\bibliography{references.bib, DLConstellationLearning.bib}

\end{document}